\documentclass[twocolumn,superscriptaddress,preprintnumbers,amsmath,amssymb]{revtex4}
\usepackage{graphicx}% Include figure files
\usepackage{dcolumn}% Align table columns on decimal point
\usepackage{bm}% bold math
\usepackage{color}
\usepackage{ulem}
\usepackage{amsmath,amssymb}
\begin{document}

\title{Relationship between the size of camphor-driven rotor and its angular velocity}

\author{Yuki~Koyano\footnote{Corresponding author. E-mail: y.koyano@chiba-u.jp.}}
\affiliation{Department of Physics, Chiba University, Chiba 263-8522, Japan}

\author{Marian~Gryciuk}
\affiliation{Institute of Physical Chemistry, Polish Academy of Sciences, Warsaw 01-224, Poland}

\author{Paulina~Skrobanska}
\affiliation{Institute of Physical Chemistry, Polish Academy of Sciences, Warsaw 01-224, Poland}

\author{Maciej~Malecki}
\affiliation{Institute of Physical Chemistry, Polish Academy of Sciences, Warsaw 01-224, Poland}

\author{Yutaka~Sumino}
\affiliation{Department of Applied Physics, Faculty of Science, Tokyo University of Science, Tokyo 125-8585, Japan}

\author{Hiroyuki~Kitahata}
\affiliation{Department of Physics, Chiba University, Chiba 263-8522, Japan}

\author{Jerzy~Gorecki}
\affiliation{Institute of Physical Chemistry, Polish Academy of Sciences, Warsaw 01-224, Poland}

\begin{abstract}
We consider a rotor made of two camphor disks glued below the ends of a plastic stripe. The disks are floating on a water surface and the plastic stripe does not touch the surface. The system can rotate around a vertical axis located at the center of the stripe. The disks dissipate camphor molecules. The driving momentum comes from the nonuniformity of surface tension resulting from inhomogeneous surface concentration of camphor molecules around the disks. We investigate the stationary angular velocity as a function of rotor radius $\ell$. For large $\ell$ the angular velocity decreases for increasing $\ell$. At a specific value of $\ell$ the angular velocity reaches its maximum and, for short $\ell$ it rapidly decreases. Such behaviour is confirmed by a simple numerical model. The model also predicts that there is a critical rotor size below which it does not rotate. Within the introduced model we analyze the type of this bifurcation. 
\end{abstract}

\maketitle

\section{Introduction}

Self-propelled particles have been intensively studied, because they can be considered as simple examples of systems imitating motion of living matter. The motion of these particles obeys laws of physics, especially with regards to symmetric properties of systems. Therefore, it is important to consider how the symmetry of self-propelled particles influences the character of their motion, because it can help in fundamental understanding of the behavior of living organisms.

Particle motion is characterized by a direction that breaks the symmetry of space.
The ability to select the direction of motion can be either embedded innately 
or acquired through spontaneous symmetry breaking.
Janus particles are an example of the first case. Particles with different surface properties between one hemisphere and the other, can move due to coupling between the surface properties and chemical reactions~\cite{Janus,Janus2}. The direction of their motion is determined by location of the reactive surface. An oil droplet containing surfactant is an example of the second type. Such a droplet can move by diffusing surfactant into the surrounding media and its direction of motion is determined by initial fluctuations~\cite{Toyota,Yoshinaga}.

The two examples mentioned above illustrate the translational motion. The same idea applies for rotational motion, as well. The rotational motion is relevant as it can occur even in a confined geometry~\cite{Camley,NakataLangmuir,Tanaka}. In order to realize the rotational motion, we also have two strategies, i.e., asymmetry embedded into the system and spontaneous symmetry breaking. The rotational motion due to the embedded asymmetry has been described for several systems~\cite{NakataLangmuir,bacteria,Mitsumata,Hayakawa,Lowen}. For example, chiral-shaped materials under the laser irradiation exhibits rotational motion~\cite{Lowen}. On the other hand, the appearance of rotational motion through spontaneous symmetry breaking is not so simple as the translational motion. There are several interesting systems that can rotate spontaneously~\cite{Bassik,Pimienta,Takabatake,Nagai}. The important work was done by Pimienta et al., who discovered spontaneous rotation of a dichloromethane droplet on water~\cite{Pimienta}. Another interesting result was reported by Takabatake et al., who demonstrated that a droplet with a small soap fragment can perform rotational motion~\cite{Takabatake,Nagai}. From these studies, we can guess that the deformation from a circular shape, i.e. an axial anisotropy, is important to realize the rotational motion through spontaneous symmetry breaking. However, a  mathematical modeling of such  motion has not been developed yet, although some preliminary attempts have been made~\cite{Pimienta2,Camley,KoyanoJCP}.

Camphor particles moving on water are good candidates to connect the experiment with mathematical model. A camphor boat and a camphor disk moving on a water surface have been intensively studied as examples of motion caused by embedded asymmetry and by spontaneous symmetry breaking, respectively~\cite{Tomlinson,Rayleigh,NakataLangmuir,pccp}. A camphor particle attached to a plastic plate moves in the direction opposite to the camphor-attached side, which is considered as a self-propelled particle with embedded asymmetry. On the other hand, a symmetric camphor disk can move in a certain direction that is determined by the initial condition or fluctuations, and it can be regarded as a self-propelled particle without embedded asymmetry. Camphor particles have also been used to study various kinds of behavior of self-propelled particles such as jamming~\cite{jamming,jamming2}, cooperative motion~\cite{Soh, Soh2, SuematsuJPSJ}, and nonequilibrium distribution of velocities~\cite{Schulz}. The motion caused by the spontaneous symmetry breaking is described in terms of bifurcations. It has been shown that the mathematical model based on the reaction-diffusion system in which the camphor surface concentration is coupled with the Newtonian equation for motion of the camphor particle reproduces such spontaneous symmetry breaking, which can be described as the pitchfork bifurcation~\cite{pccp,Nagayama}.

The rotational motion due to the embedded asymmetry has been previously observed also in camphor systems. For example, comma-shaped camphor particles~\cite{pccp, NakataLangmuir} and a propeller made of camphor disks and plastic plates~\cite{Nakatamobile} exhibit rotational motion in the direction determined by the asymmetry. In order to realize the rotational motion through spontaneous symmetry breaking, we have to introduce an axial anisotropy. Since camphor is solid, it is easy to introduce anisotropy by designing non-symmetric particles. One of the authors (H.K.) has analytically demonstrated a rotational motion through spontaneous symmetry breaking using an elliptic camphor disk~\cite{IKN}. It should be noticed that elliptic shape has chiral symmetry and thus the rotational motion of an elliptic camphor particle appears due to the spontaneous symmetry breaking between clockwise and anticlockwise rotational modes. Though a mathematical description of elliptic camphor system is relatively simple, there are still difficulties in the analytical approach. Therefore, we consider a simpler system that has anisotropy but keeps chiral symmetry, and can exhibit rotational motion through spontaneous symmetry breaking. In this paper, we consider a camphor rotor, which is composed of two camphor disks rigidly interconnected with each other (cf. Fig.~\ref{fig1}). For such a rotor, the character of bifurcation can be analytically investigated.

\section{Experiments}

\begin{figure}
	\begin{center}
		\includegraphics[width=85mm]{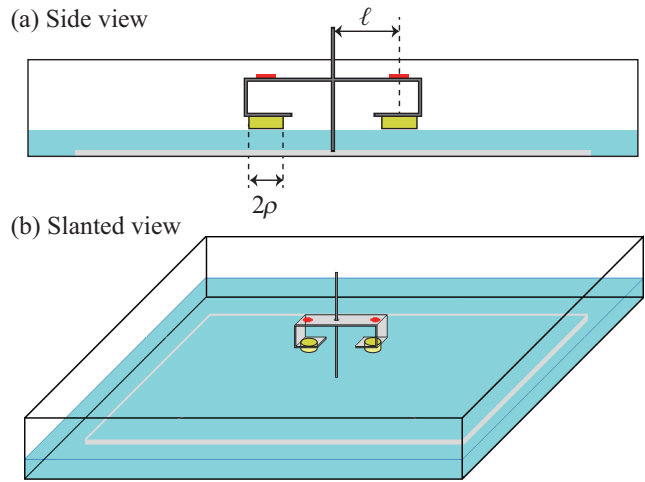}
		\caption{Schematic illustration of the experimental setup. The side view (a) and the slanted view (b) of the rotor are shown.}
		\label{fig1}
	\end{center}
\end{figure}

\begin{figure}
	\begin{center}
		\includegraphics[width=85mm]{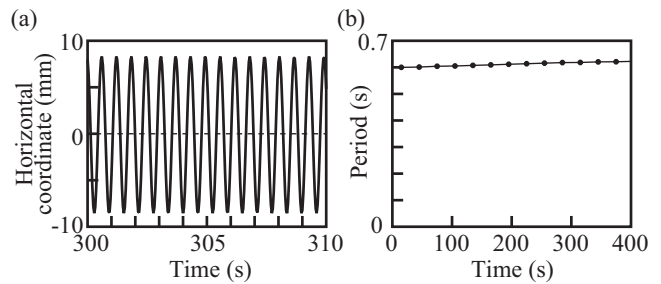}
		\caption{Experimental results on rotor motion. (a) The time evolution of a horizontal coordinate of one of marking dots for a rotor with $\ell= 8.5$~mm in the time interval from $300$~s to $310$~s.	(b) The period for the rotor with $\ell= 8.5$~mm as the function of time.}
		\label{fig2}
	\end{center}
\end{figure}

We study the motion of a simple rotor powered by two camphor disks glued below the ends of a plastic stripe as illustrated in Fig.~\ref{fig1}. The system can rotate around a vertical axis located at the center of the stripe. Commercially available camphor ($ 99 \%$ purity, Sigma-Aldrich) was used without further purification. The disks were made by pressing camphor in a pill maker. The radius of each camphor disk was $\rho = 1.5$~mm and it was $1$~mm high. The rotor was floating on a water surface in the square tank (tank side $120$~mm) and the water level was $10$~mm. In order to reduce the hydrodynamic flows the central part of the plastic stripe was elevated above the water level so that only the bottom surface of camphor disks had contact with water and the stripe did not touch its surface. The profile of camphor surface concentration on water results from the balance between the inflow of camphor molecules from the disks and camphor evaporation into the air and dissolution in the water~\cite{pccp}. It is known that water surface tension is a decreasing function of camphor surface concentration~\cite{NakataLangmuir,pccp,Suematsu}. The averaged force acting on a camphor disk is directed towards the region with the lowest camphor surface concentration around the disk. The driving torque of the rotor comes from differences of surface tension around the disks resulting from inhomogeneous surface concentration of camphor molecules. The time evolution of rotor was recorded using a digital camera (NEX VG20EH, SONY) and the coordinates of red dots (cf. Fig.~\ref{fig2}b) located over the centers of camphor disks were obtained using the ImageJ software~\cite{ImageJ}. A typical time of experiment was in the range from 5 to 10 minutes. 

The distance between the axis and the disk center $2 \ell$ was the control parameter for our experiments. Periodic changes in the horizontal coordinate of one of the dots for the rotor with $\ell = 8.5$~mm are shown in Fig.~\ref{fig2}a. During the time of all experiments we observed highly regular rotations without any significant perturbations of rotor motion. The period of oscillations was measured as the time between the successive maxima separately in each $30$~s interval. Typically the period slowly increased with time as illustrated in Fig.~\ref{fig2}b. The changes were not significant and for the subsequent analysis we consider the values obtained in the time interval from $300$~s to $400$~s.

\begin{figure}
	\begin{center}
		\includegraphics[width=85mm]{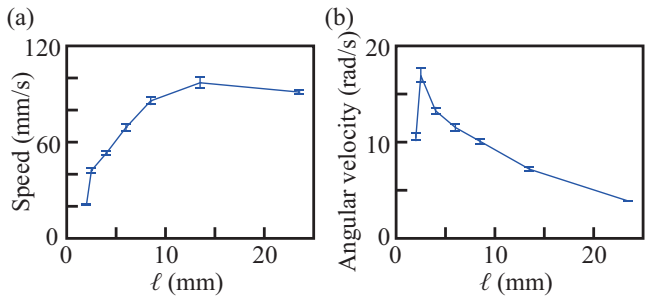}
		\caption{Experimental results on rotor motion as a function of rotor radius $\ell$. (a) The speed of the center of camphor disk. (b) The angular velocity of a rotor. The bars estimate the experimental errors.}
		\label{fig3}
	\end{center}
\end{figure}

Fig.~\ref{fig3} illustrates the speed of disks center (a) and the angular velocity (b) as the function of $\ell$. The speed grew monotonically with $\ell$. These results can be explained by the larger radius of motion. It can be expected that for large $\ell$ the speed saturates to be the one for a separated camphor disk. An interesting behaviour was observed for the angular velocity as a function of $\ell$. For large $\ell$ it was a decreasing function. It reached its maximum around $\ell=2.5$~mm and then rapidly dropped. In the following sections we present  numerical and theoretical arguments explaining such behaviour and study the type of bifurcation leading to rotations above the critical value of rotor radius. 

\section{The Mathematical Model}

In order to discuss the mechanisms of rotor motion, we consider a mathematical model presented below. We define the center position of the $i$-th camphor disk as $\bm{\ell}_i(t)$. The center of mass of both camphor disks is fixed to the origin of coordinate system ($(\bm{\ell}_1(t) + \bm{\ell}_2(t)) / 2 = \bm{0}$). Thus, the positions of camphor disks center can be defined only using a single angle $\theta(t)$, i.e.,
\begin{align}
\bm{\ell}_1 (t) = \ell \bm{e} (\theta (t)), \quad 
\bm{\ell}_2 (t) = -\ell \bm{e} (\theta (t)), \label{def.l}
\end{align}
where we set a unit vector $\bm{e} (\theta(t))$ as $\bm{e} (\theta) = \bm{e}_x \cos \theta + \bm{e}_y \sin \theta$, and $\bm{e}_x$ and $\bm{e}_y$ are the unit vectors along the $x$- and $y$- axes, respectively. 

The time evolution of the surface concentration field of camphor molecules $c(\bm{r}, t)$ is described as~\cite{pccp,Nagayama}
\begin{equation}
\frac{\partial c}{\partial t} = \nabla^2 c - c + f, \label{conc.eq.}
\end{equation} 
where $-c$ describes sublimation and dissolution of camphor molecules and $f = f(\bm{r}; \bm{\ell}_1, \bm{\ell}_2)$ is a function representing the supply of camphor molecules from the camphor disks.
Equation~\eqref{conc.eq.} is written using dimensionless variables. The real length, time, and concentration are normalized with the diffusion length $\sqrt{D / a}$, the characteristic time of sublimation/dissolution $1/a$, and the ratio between the supply and dissipation rates of camphor, $f_0/a$, where $a$ is the dissipation rate of camphor, $f_0$ is the total inflow of camphor from a single disk per unit of real time, and $D$ is the effective diffusion constant of camphor molecules.
Experimental observations show that the effective diffusion of camphor molecules is much faster than the thermodynamical one~\cite{Suematsu}.
It has also been also experimentally confirmed that the diffusion enhancement results from the Marangoni flow in aqueous phase.
The enhancement of diffusion by Marangoni flow was analytically described by the model based on the Stokes equation coupled with the equation for camphor surface concentration dissolving from a fixed disk~\cite{Kitahata_arxiv}. Following this result, we assumed that also in the case of rotor the hydrodynamic effects can be approximated by the effective diffusion constant, and we do not explicitly include hydrodynamics in our model.
This assumption is valid when camphor disks slowly move. Thus it can be used to analyze analytically the transition between the stationary and moving rotor as the function of rotor radius.

Time evolution of $\theta(t)$ is described as
\begin{equation}
I (\ell) \frac{d^2 \theta}{dt^2} = - \eta (\ell) \frac{d\theta}{dt} + \mathcal{T}, \label{eq-theta}
\end{equation}
where $I$ and $\eta$ are the moment of inertia and the friction coefficient of the camphor disks, respectively, and they depend on $\ell$ as follows:
\begin{align}
I (\ell) =& 2 \pi \rho^2 \sigma \ell^2, \label{def.I} \\
\eta (\ell) =& 2 \pi \rho^2 \kappa \ell^2, \label{def.eta}
\end{align}
where $\sigma$ and $\kappa$ are dimensionless parameters corresponding to the mass and the friction constant per unit area for the camphor disks, respectively. Here, the friction force working on the $i$-th camphor disk is described as $- (\pi \rho^2 \kappa) \dot{\bm{\ell}}_i$.
The detailed derivation is presented in Appendix~\ref{dev.eq-theta}.

In Eq.~\eqref{eq-theta}, $\mathcal{T}$ is the torque with respect to the origin acting on the rotor: \begin{equation}
\mathcal{T} = \sum_{i=1}^2 \bm{\ell}_i \times \left[ \int_0^{2\pi} \gamma \left( c\left( \bm{\ell}_i +  \rho \bm{e}(\phi) \right)\right) \bm{e}(\phi) \rho d\phi \right], \label{torque1}
\end{equation}
where $\gamma (c)$ is a function that represents the dependence of the surface tension on the surface concentration of camphor molecules. 
Here, we define the operator ``$\times$'' as:
\begin{align}
{\bm a} \times {\bm b} = a_1 b_2 - a_2 b_1
\end{align}
for two dimensional vectors ${\bm a} = a_1 {\bm e}_x + a_2 {\bm e}_y$, and ${\bm b} = b_1 {\bm e}_x + b_2 {\bm m}_y$.
If we assume that the surface tension $\gamma$ is a linear decreasing function of $c$, i.e.,
\begin{equation}
\gamma(c) = \gamma_0 - k c, 
\end{equation}
where $\gamma_0$ is the surface tension of pure water, and $k$ is a positive constant, then Eq.~\eqref{torque1} can be rewritten as
\begin{align}
\mathcal{T} = - k \ell \bm{e} (\theta)
& \times \left[ \int_0^{2\pi} c\left( \bm{\ell}_1 +  \rho \bm{e}(\phi) \right) \bm{e}(\phi) \rho d\phi \right. \nonumber \\
& \qquad \left. - \int_0^{2\pi} c\left( \bm{\ell}_2 +  \rho \bm{e}(\phi) \right) \bm{e}(\phi) \rho d\phi \right]. \label{torque2}
\end{align}
Hereafter, we set $k=1$ without losing generality.

\section{Numerical Simulations of Camphor-Driven rotor}

\begin{figure}
	\begin{center}
		\includegraphics{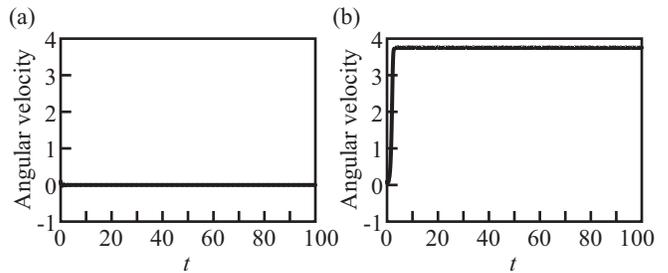}
	\end{center}
	\caption{Angular velocity as a function of time for a small and a large rotor: (a) $\ell = 0.3$  and (b) $\ell = 0.5$. The initial conditions were $\theta = 1$, $d\theta/ dt = 0.1$, and $c = 0$ at all space points.}
	\label{fig4}
\end{figure}

\begin{figure}
	\begin{center}
		\includegraphics{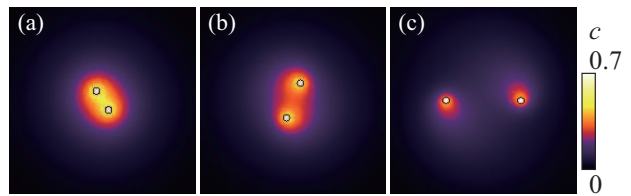}
	\end{center}
	\caption{Profiles of camphor concentration at $t = 100$ for (a) $\ell = 0.3$, (b) $\ell = 0.5$, and (c) $\ell = 1.0$. The rotor does not move in (a) and it rotates clockwise in (b) and (c). The initial conditions were all the same as those in Fig.~\ref{fig4}.}
	\label{fig5}
\end{figure}

We performed numerical simulations of the rotor dynamics according to Eqs.~\eqref{conc.eq.} and \eqref{eq-theta}.
The supply rate from the camphor disk in Eq.~\eqref{conc.eq.} is given as
\begin{equation}
f(\bm{r}; \bm{\ell}_1, \bm{\ell}_2) = \sum_{i = 1,2} \frac{1}{\pi \rho^2} \left [ \frac{1}{2} \left( 1 + \tanh \frac{\rho - \left| \bm{r} - \bm{\ell}_i \right|}{\delta}  \right) \right ],
\end{equation}
where $\delta$ is a smoothing parameter set to be $\delta = 0.025$.
The total supply from a single camphor disk is approximately equal to $1$.
We used the Euler method to calculate the reaction terms, and explicit method for the diffusion. 
The time step was $\Delta t = 10^{-4} $ and the spatial step was $\Delta x = 0.025$. 
We set parameters as $\rho = 0.1$, $\sigma = 0.004$, and $\kappa = 0.12$.
As for the concentration field, we consider a circular outer boundary with a radius of $10$, which hardly affects the motion of the rotor for $\ell \leq 5$.
In order to calculate the force acting on each camphor disk in Eq.~\eqref{torque2}, we replaced the integration in Eq.~\eqref{torque2} into the summation over 32 arc elements. 
We performed numerical simulations and obtained the time evolution of the angle $\theta(t)$ and the angular velocity $d\theta/dt$. 
The distance between two camphor disks, $\ell$, controls the behaviour of a rotor.
For larger $\ell$, the rotor moves stationarily, whereas for smaller $\ell$, it stops as shown in Fig.~\ref{fig4}.
The snapshots for the camphor concentration for various $\ell$ are shown in Fig.~\ref{fig5}. 
In the case when the rotor does not move, the camphor concentration profile is symmetric with respect to the axis connecting the centers of two camphor disks as in Fig.~\ref{fig5}a. In contrast, if it rotates, the profile has chiral asymmetry as shown in Fig.~\ref{fig5}b and c.  

\begin{figure}
	\begin{center}
		\includegraphics{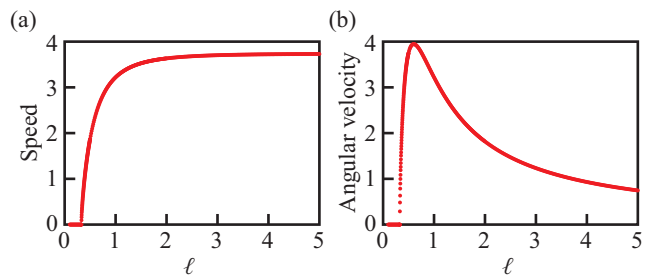}
	\end{center}
	\caption{Numerical results on stationary speed (a) and stationary angular velocity (b) as a function of the rotor radius $\ell$.
	}
	\label{fig6}
\end{figure}

In Fig.~\ref{fig6}, we present the stationary speed of disks center and the stationary angular velocity of rotor as a function of rotor radius $\ell$. 
For the large $\ell$, we expect that the interactions between the two camphor disks becomes negligible. In such a case, the both camphor disks should move at the speed equal to that for a single camphor disk without any constraints. Then, the angular velocity should be inversely proportional to $\ell$.
For small $\ell$, we can see the transition-like behaviour between static and moving rotor around $\ell \simeq 0.33$ in Fig.~\ref{fig6}. We expect this transition originates from pitchfork bifurcation, at which the stable rest state becomes unstable.

\section{Analysis of critical slowing down}

In this section, the dynamical system for the angular velocity of a single rotor is derived by the reduction of the model equations and its bifurcation structure is revealed.
We consider the limit of $\rho \to +0$, i.e., the case where the camphor disks radius is small enough compared with the diffusion length ($=1$) and the radius of rotor ($=\ell$).

By dividing the both sides of Eq.~\eqref{eq-theta} with $\pi \rho^2 \ell^2$, we obtain
\begin{equation}
\sigma \frac{d^2 \theta}{dt^2} = - \kappa \frac{d\theta}{dt} + \frac{1}{2 \pi \rho^2 \ell^2} \mathcal{T}. \label{/2pirho2ell2}
\end{equation}
Here, we take in the limit of $\rho \to +0$, and we obtain
\begin{equation}
\frac{1}{\pi \rho^2} \mathcal{T} \; \to \lim_{\rho \to +0} \frac{1}{\pi \rho^2} \mathcal{T} = - \sum_{i = 1,2} \bm{\ell}_i \times \left . \nabla c (\bm{r}) \right |_{\bm{r}=\bm{\ell}_i} ,\label{torque3}
\end{equation}
from the simple calculation for the concentration $c(\bm{r})$ with no divergence at $\bm{r} = \bm{\ell}_i$.

The equation for the concentration field is represented in Eq.~\eqref{conc.eq.}.
The source term $f$ in Eq.~\eqref{conc.eq.} is given by
\begin{equation}
f(\bm{r}; \bm{\ell}_1, \bm{\ell}_2) = \sum_{i=1,2} \delta (\bm{r}-\bm{\ell}_i) = \sum_{i=1,2} \frac{1}{r} \delta (r-\ell) \delta (\phi-\theta_i),
\end{equation}
since we consider that the size of the camphor disks is infinitesimally small.
Here, $\bm r$ is represented as $\bm{r} = (r,\phi)$ in the polar coordinates.

The concentration field is the summation of the concentration field made by each camphor disk since the equation for the concentration field is linear.
Thus, the concentration field made by a rotor is given by
\begin{align}
c(\bm{r}) = c_s(\bm{r}; \bm{\ell}_1) + c_s(\bm{r}; \bm{\ell}_2), \label{c=cs+cs}
\end{align}
where $c_s(\bm{r}; \bm{\ell})$ is the concentration field made by a single camphor disk located at $\bm{\ell}$, i.e., the solution of Eq.~\eqref{conc.eq.} with the source term $\delta (\bm{r}-\bm{\ell})$.
When the velocity of the camphor disk is sufficiently small, the concentration field made by a single disk $c_s(\bm{r}; \bm{\ell})$ is analytically expressed as
\begin{align}
c_s(\bm{r}; \bm{\ell}) =& c_{00} (\lambda) + c_{10} (\lambda) (\bm{r} - \bm{\ell}) \cdot \dot{\bm{\ell}} \nonumber \\
 &+ c_{20} (\lambda) (\bm{r} - \bm{\ell}) \cdot \ddot{\bm{\ell}} + c_{21}(\lambda) \left| \dot{\bm{\ell}} \right|^2 \nonumber \\ & + c_{22}(\lambda) \left[(\bm{r} - \bm{\ell}) \cdot \dot{\bm{\ell}}  \right]^2 \nonumber \\
&+ c_{30} (\lambda) (\bm{r} - \bm{\ell}) \cdot \dddot{\bm{\ell}} + c_{31}(\lambda) \left| \dot{\bm{\ell}} \right|^2 (\bm{r} - \bm{\ell}) \cdot \dot{\bm{\ell}}\nonumber \\
&+ c_{32}(\lambda) \left[(\bm{r} - \bm{\ell}) \cdot \dot{\bm{\ell}}  \right]^3 + c_{33}(\lambda) \dot{\bm{\ell}} \cdot \ddot{\bm{\ell}}\nonumber\\
&+ c_{34}(\lambda) \left[(\bm{r} - \bm{\ell}) \cdot \dot{\bm{\ell}}  \right] \left[(\bm{r} - \bm{\ell}) \cdot \ddot{\bm{\ell}}  \right],
\end{align}
where $\lambda = \left| \bm{r} - \bm{\ell} \right|$, the dot over variables ($\; \dot{} \;$) represents the time derivative, and the dot between vectors ($\cdot$) represents inner product. Here,
\begin{align}
& c_{00}(\lambda) = \frac{1}{2\pi} \mathcal{K}_0 \left ( \lambda \right ), && c_{10}(\lambda) = - \frac{1}{4\pi} \mathcal{K}_0 \left ( \lambda \right ), \nonumber \\
& c_{20}(\lambda) = \frac{1}{16\pi} \lambda \mathcal{K}_1 \left (\lambda \right ), && c_{21}(\lambda) =- \frac{1}{16\pi} \lambda \mathcal{K}_1 \left (\lambda \right ), \nonumber \\
& c_{22}(\lambda) = \frac{1}{16\pi} \mathcal{K}_0 \left ( \lambda \right ), &&  c_{30}(\lambda) = -\frac{1}{96\pi} \lambda^2 \mathcal{K}_2(\lambda), \nonumber \\
& c_{31}(\lambda) = \frac{1}{32\pi} \lambda \mathcal{K}_1  \left (\lambda \right ), && c_{32}(\lambda) = -\frac{1}{96\pi} \mathcal{K}_0(\lambda), \nonumber \\
&c_{33}(\lambda) = \frac{1}{32\pi} \lambda^2 \mathcal{K}_2(\lambda), && c_{34}(\lambda)= -\frac{1}{32\pi} \lambda \mathcal{K}_1(\lambda), \label{conc_single}
\end{align}
where $\mathcal{K}_n$ is the second-kind modified Bessel function of the $n$-th order.
It is noted that the term composed of variables with totally more-than-three-time derivatives is neglected.
The derivation is shown in Appendix~\ref{conc.field}.

From Eq.~\eqref{c=cs+cs}, the torque per contact area \eqref{torque3} is represented as
\begin{equation}
\lim_{\rho \to +0} \frac{1}{\pi \rho^2} \mathcal{T} = \sum_{i,j = 1,2} \tau_{ij}.
\end{equation}
$\tau_{ii}$ is the torque per contact area working on a camphor disk originating from self-made concentration field, and calculated as 
\begin{align}
\tau_{ii} =& \bm{\ell}_i \times \lim_{\rho \to +0} \frac{-1}{\pi \rho^2} \int_0^{2\pi} c_s \left( \bm{\ell}_i +  \rho \bm{e}(\phi) ; \bm{\ell}_j \right) \bm{e}(\phi) \rho d\phi \nonumber \\
=& \frac{1}{4\pi} \left ( - \gamma_{\mathrm{Euler}} + \log \frac{2}{\rho} \right ) \ell^2 \dot{\theta} - \frac{1}{16\pi} \ell^2 \ddot{\theta} - \frac{1}{32\pi} \ell^4 \dot{\theta}^3 \nonumber \\
& + \frac{1}{48 \pi} \ell^2 \left(\dddot{\theta} - \dot{\theta}^3 \right), \label{torqueii}
\end{align}
where $\gamma_{\mathrm{Euler}}$ is Euler constant ($\simeq 0.577$).
Here we used $\bm{\ell} \times \dot{\bm{\ell}} = \ell^2 \dot{\theta}$, $\bm{\ell} \times \ddot{\bm{\ell}} = \ell^2 \ddot{\theta}$, and $\bm{\ell} \times \dddot{\bm{\ell}} = \ell^2 \left ( \dddot{\theta} - \dot{\theta}^3 \right )$. 

Then, we consider the torque working on one camphor disk by the other camphor disk.
Since the concentration field $c_s(\bm{r}; \bm{\ell})$ does not diverge except at $\bm{r} = \bm{\ell}$, the torque per contact area by the other camphor disk, $\tau_{ij}$ $(i \neq j)$, is calculated as
\begin{align}
\tau_{ij} =& -\frac{1}{4\pi} \mathcal{K}_0 \left ( 2 \ell \right ) \ell^2 \dot{\theta} + \frac{1}{8\pi} \mathcal{K}_1 \left ( 2 \ell \right ) \ell^3 \ddot{\theta} \nonumber \\
& - \frac{1}{16\pi} \mathcal{K}_1 \left ( 2 \ell \right ) \ell^5 \dot{\theta}^3 - \frac{1}{24 \pi} \mathcal{K}_2 (2 \ell) \ell^4 \left ( \dddot{\theta} - \dot{\theta}^3 \right ), \label{torque4}
\end{align}
by using Eq.~\eqref{torque3}.

From Eqs.~\eqref{/2pirho2ell2}, \eqref{torqueii} and \eqref{torque4}, we have the reduced equation:
\begin{align}
\sigma \ddot{\theta} =& - \kappa \dot{\theta} + \frac{1}{2 \ell^2} \sum_{i,j = 1,2} \tau_{ij} \\
=& - \kappa \dot{\theta} + \frac{1}{4\pi} \left ( -\gamma_{\mathrm{Euler}} +\log \frac{2}{\rho} - \mathcal{K}_0 \left ( 2 \ell \right ) \right ) \dot{\theta} \nonumber \\
& - \frac{1}{16\pi} \left ( 1 - 2 \ell \mathcal{K}_1 \left ( 2 \ell \right ) \right ) \ddot{\theta} \nonumber \\
& - \frac{1}{32\pi} \left ( 1 + 2 \ell \mathcal{K}_1 \left ( 2 \ell \right ) \right ) \ell^2 {\dot{\theta}}^3 \nonumber \\
& + \frac{1}{48 \pi} \left ( 1 - 2 \ell^2 \mathcal{K}_2 (2 \ell) \right ) \left ( \dddot{\theta} - \dot{\theta}^3 \right ) .
\end{align}

\begin{figure}
	\begin{center}
		\includegraphics{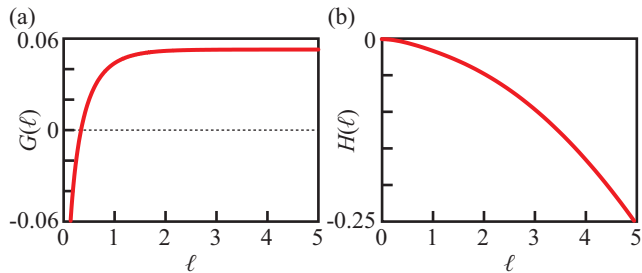}
		\caption{Coefficients $G(\ell)$ and $H(\ell)$. The parameters are set to be $\kappa = 1.2$ and $\rho = 0.1 e^{1/4}$~\cite{note}.}
		\label{fig-BK}
	\end{center}
\end{figure}

Based on the description, we discuss a bifurcation structure.
We consider the stable solution of $\dot{\theta} = \mathrm{const}. \equiv \omega$.
When the rotor rotates with a constant angular velocity, $\dot{\omega}$ and $\ddot{\omega}$ should be zero.
Thus we have
\begin{align}
&\left [ \frac{1}{4\pi} \left ( -\gamma_{\mathrm{Euler}} + \log \frac{2}{\rho} - \mathcal{K}_0 \left ( 2 \ell \right ) \right ) - \kappa \right ] \omega \nonumber \\
&- \frac{1}{96\pi} \left [ 3 \left ( 1 + 2 \ell \mathcal{K}_1 \left ( 2 \ell \right ) \right ) \ell^2 + 2 \left ( 1 - 2 \ell^2 \mathcal{K}_2 (2 \ell) \right ) \right ] \omega^3 = 0.
\end{align}
Here, we define the coefficients of $\omega$ and $\omega^3$ as $G(\ell) = \left [ -\gamma_{\mathrm{Euler}} +\log (2/\rho) - \mathcal{K}_0 \left ( 2 \ell \right ) \right ] / (4\pi) - \kappa$ and $H(\ell) = - \left [ 3 \left ( 1 + 2 \ell \mathcal{K}_1 \left ( 2 \ell \right ) \right ) \ell^2 + 2 \left ( 1 - 2 \ell^2 \mathcal{K}_2 (2 \ell) \right ) \right ] /(96 \pi)$, respectively.
The dependences of $G(\ell)$ and $H(\ell)$ on $\ell$ are displayed in Fig.~\ref{fig-BK}.
The stable angular velocity is realized when $G(\ell)$ is positive and $H(\ell)$ is negative, and thus the bifurcation point is $\ell = \ell_c$, where $G(\ell_c) = 0$.

\begin{figure}
	\begin{center}
		\includegraphics{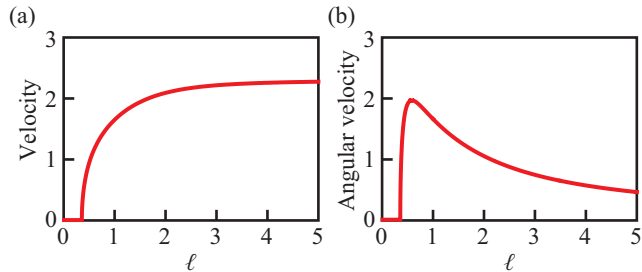}
		\caption{Velocity and angular velocity depending on $\ell$. Parameters are $\kappa = 1.2$ and $\rho = 0.1 e^{1/4}$~\cite{note}.}
		\label{fig-phidot}
	\end{center}
\end{figure}

The stable angular velocity $\omega$ is given by $\sqrt{-G(\ell)/H(\ell)}$ for $G(\ell) > 0$ and $0$ for $G(\ell) < 0$, and its dependence on $\ell$ is shown in Fig.~\ref{fig-phidot}. At $\ell \simeq 0.35$, pitchfork bifurcation occurs when we set the parameters as $\kappa = 1.2$ and $\rho = 0.1 e^{1/4}$~\cite{note}. Over the bifurcation point, the rest state becomes unstable and rotational motion occurs with a constant angular velocity. The bifurcation point is quantitatively corresponding to the numerical results shown in Fig.~\ref{fig6}. 

\section{Discussion}

In the present paper, we discuss the dependence between the angular velocity of a camphor-driven rotor and its radius. Experiments have demonstrated that the angular velocity increases with decreasing radius, reaches its maximum and then rapidly falls. 
We have introduced a mathematical model in which the rotor motion is coupled with the surface concentration of camphor via concentration-dependent surface tension under the assumption of slow velocity of camphor disks. 
The model predicts that rotors with smaller radius do not rotate because the surface concentration of camphor becomes symmetric with respect to the line connecting disk centers and the torque acting on the rotor vanishes. Simulations demonstrate a sharp transition between rotating and non-rotating states at a specific rotor radius. Considering the rotor radius as a control parameter, we have examined the bifurcation between rotating and non-rotating states. The evolution equations have been simplified assuming that camphor disks powering the rotor are infinitesimally small. The simplified model has been solved analytically and the formula for the critical rotor radius is derived. Moreover, we have shown that the bifurcations between the rotating and non-rotating states is of the pitchfork type. 

Although hydrodynamics is not explicitly included in our model,
the dependence of angular velocity on rotor radii is qualitatively the same as the experimental one.
We can speculate that, even though camphor disk velocity is not small, the hydrodynamic effects can be still taken into account by the effective diffusion of camphor in Eq.~\eqref{conc.eq.}.
The detailed analysis of such approximation is planed for the future studies.

We believe that the methods presented in the paper can inspire further studies on complex behaviour in other systems powered by changes of surface tension, like, for example, interactions between multiple camphor-powered rotors.
The synchronization between such rotors has been observed experimentally. We expect to find various kinds of interesting behaviors originating from cooperativity in the system with spatially distributed rotors.

\begin{acknowledgments}

The authors acknowledge Professor S.~Nakata for his helpful discussion.
This work was supported by JSPS-PAN Bilateral Joint Research Program ``Spontaneous creation of chemical computing structures based on interfacial interactions'' between Japan and the Polish Academy of Sciences, JSPS KAKENHI Grants No.~JP16K13866 and JP16H06478 to Y.S., No.~JP25103008, JP15K05199, and JP16H03949 to H.K., the Sasakawa Scientific Research Grant from the Japan Science Society to Y.K. (No.~28-225), and the Cooperative Research Program of ``Network Joint Research Center for Materials and Devices'' No.~20165001 and 20175002 to Y.K., No.~20161033 and 20171033 to Y.S., and No.~20163002 and 20173006 to H.K.

\end{acknowledgments}

\appendix

\section{Derivation of Eq.~\eqref{eq-theta}\label{dev.eq-theta}}

Here, we consider the equation of motion of a rotor.
The equation of motion for each camphor disk is represented as
\begin{align}
\pi \rho^2 \sigma \ddot{\bm{\ell}}_1 = - \pi \rho^2 \kappa \dot{\bm{\ell}}_1 + \bm{F}_1, \label{eq1} \\
\pi \rho^2 \sigma \ddot{\bm{\ell}}_2 = - \pi \rho^2 \kappa \dot{\bm{\ell}}_2 + \bm{F}_2. \label{eq2}
\end{align}
It is noted that $\pi \rho^2 \sigma$ and $\pi \rho^2 \kappa$ are the mass and friction constant of a camphor disk, respectively.
The definition of $\bm{F}_1$ and $\bm{F}_2$ is 
\begin{align}
\bm{F}_1 = \bm{F} (c;\bm{\ell}_1) + \bm{F}_{\rm constraint}^{(1)}, \\
\bm{F}_2 = \bm{F} (c;\bm{\ell}_2) + \bm{F}_{\rm constraint}^{(2)}, 
\end{align}
where
\begin{align}
\bm{F} (c;\bm{\ell}_i) = \int_0^{2\pi} \gamma \left( c\left( \bm{\ell}_i +  \rho \bm{e}(\phi) \right)\right) \bm{e}(\phi) \rho d\phi.
\end{align}
Here, $\bm{F}_{\rm constraint}^{(i)}$ is given by $\bm{F}_{\rm constraint}^{(i)} = - [\bm{F} (c;\bm{\ell}_i) \cdot \bm{\ell}_i ] \bm{\ell}_i / \ell^2$, which means that the constraint force balances the component of driving force proportional to $\bm{\ell}_i$.

By taking the vector product of the both sides of equation of motion Eqs.~\eqref{eq1} and \eqref{eq2} with $\bm{\ell}_i$ and taking summation of them, we obtain, 
\begin{align}
I(\ell) \ddot{\theta} = - \eta (\ell) \dot{\theta} + \mathcal{T},
\end{align}
where $I(\ell) = 2 \pi \rho^2 \sigma \ell^2$ and $\eta(\ell) = 2 \pi \rho^2 \kappa \ell^2$, and the definition of the torque $\mathcal{T}$ is given in Eq.~\eqref{torque1}.
Here we used $\dot{\theta}_1 = \dot{\theta}_2$ and $\ddot{\theta}_1 = \ddot{\theta}_2$.

\section{Concentration Field of Camphor Molecules \label{conc.field}}

We consider the concentration field made by a moving camphor disk.
The concentration field and source term is expanded as follows.
\begin{align}
c_s(\bm{r}; \bm{\rho}) =& \frac{1}{2 \pi} \sum_{m = -\infty}^{\infty} \int_{0}^{\infty}  c_m(k) J_{|m|} (k r) e^{i m \phi} k dk, \nonumber \\
f(\bm{r}; \bm{\rho}) =& \frac{1}{r} \delta (r-\rho(t)) \delta (\phi - \theta(t)) \nonumber \\
=& \frac{1}{2 \pi} \sum_{m = -\infty}^{\infty} \int_{0}^{\infty} J_{|m|} (k \rho(t)) J_{|m|} (k r) e^{i m (\phi -\theta(t))}.
\end{align}
Here, the $r$- and $\phi$-directions are expanded with Hankel transform and into Fourier series, respectively.
By putting these expansion into Eq.~\eqref{conc.eq.}, we have
\begin{align}
\frac{\partial c_m(k)}{\partial t} = - (k^2 + 1) c_m(k) + J_{|m|} (k \rho(t)) e^{-i m \theta(t)}.
\end{align}
To solve the above equation, the Green's function $g_m(k,t)$ in wavenumber space, which satisfies
\begin{equation}
\frac{\partial g_m(k)}{\partial t} = - (k^2 + 1) g_m(k) + \delta(t),
\end{equation}
is obtained as
\begin{equation}
g_m(k,t) 
= e^{ -(k^2 + 1) t} \Theta (t),
\end{equation}
where $\Theta(t)$ is the Heaviside function.
The concentration field in wavenumber space, $c_m(k)$, is represented by using $g_m(k,t)$ as
\begin{align}
c_m(k,t) =& \int_{-\infty}^{\infty} J_{|m|} (k \rho({t}')) e^{-i m \theta({t}')} g_m(k,t-{t}') d{t}' \nonumber \\
=& e^{-(k^2 + 1) t} \int_{-\infty}^{t} J_{|m|} (k \rho({t}')) e^{-i m \theta({t}')} e^{ (k^2 + 1) {t}'} d{t}',
\end{align}
The above integration is expanded by partial integral~\cite{Shitara,Koyano} as follows: 
\begin{widetext}
\begin{align}
c_m(k) 
%=& \frac{e^{-At}}{A} \left [ J_{|m|} (k \rho({t}')) e^{-i m \theta({t}')} e^{ A {t}'} \right ]_{-\infty}^{t} \nonumber \\
%& - \frac{e^{-A t}}{A} \int_{-\infty}^{t} \left \{ k \dot{\rho}({t}') {J'_{|m|}} (k \rho({t}')) - i m \dot{\theta}({t}') J_{|m|} (k \rho({t}')) \right \} e^{-i m \theta({t}')} e^{ A{t}'} d{t}' \nonumber \\
%=& \cdots \nonumber \\
=& \frac{1}{A} J_{|m|} (k \rho(t)) e^{-i m \theta(t)} - \frac{1}{A^2} \left \{ k \dot{\rho}(t) {J'_{|m|}} (k \rho(t)) - i m \dot{\theta}(t) J_{|m|} (k \rho(t)) \right \} e^{-i m \theta(t)} \nonumber \\
& + \frac{1}{A^3} \left \{ k \ddot{\rho}(t) {J'_{|m|}} (k \rho(t)) + k^2 (\dot{\rho}(t))^2 {J''_{|m|}} (k \rho(t)) \right . \nonumber \\
& \qquad \quad \left . - 2 i k m \dot{\rho}(t)\dot{\theta}(t) {J'_{|m|}} (k \rho(t)) - i m \ddot{\theta}(t) J_{|m|} (k \rho(t)) - m^2 (\dot{\theta}(t))^2 J_{|m|} (k \rho(t)) \right \} e^{-i m \theta(t)} \nonumber \\
& - \frac{1}{A^4} \left \{ k^3 (\dot{\rho}(t))^3 {J'''_{|m|}} (k \rho(t)) - 3 i k^2 m (\dot{\rho}(t))^2 \dot{\theta}(t) {J''_{|m|}} (k \rho(t)) - 3 k m^2 \dot{\rho}(t) (\dot{\theta}(t))^2 {J'_{|m|}} (k \rho(t)) \right . \nonumber \\
& \qquad \quad \left .+ i m^3 (\dot{\theta}(t))^3 J_{|m|} (k \rho(t)) +k \dddot{\rho}(t) J'_{|m|}(k\rho(t)) + 3k^2 \dot{\rho}(t) \ddot{\rho}(t) J''_{|m|}(k\rho(t)) - 3ikm \ddot{\rho}(t) \dot{\theta}(t) J'_{|m|}(k\rho(t)) \right . \nonumber \\
& \qquad \quad \left .  - 3ikm \dot{\rho}(t) \ddot{\theta}(t) J'_{|m|}(k\rho(t))  - im \dddot{\theta}(t) J_{|m|}(k\rho(t)) - 3m^2 \dot{\theta}(t) \ddot{\theta}(t) J_{|m|}(k\rho(t)) \right \} e^{-i m \theta(t)} \nonumber \\
& + \cdots, \nonumber
\end{align}
where $A = k^2 + 1$. By neglecting the higher order terms, the concentration field in real space is obtained as 
\begin{align}
c_s (\bm{r}; \bm{\rho}) 
=& \frac{1}{2 \pi} \sum_{m = -\infty}^{\infty} \int_{0}^{\infty} \frac{1}{A} J_{|m|} (k \rho(t)) J_{|m|} (k r) e^{i m (\phi-\theta(t))} k dk \nonumber \\
& - \frac{1}{2 \pi} \sum_{m = -\infty}^{\infty} \int_{0}^{\infty} \frac{1}{A^2} \left \{ k \dot{\rho}(t) {J'_{|m|}} (k \rho(t)) - i m \dot{\theta}(t) J_{|m|} (k \rho(t)) \right \} J_{|m|} (k r) e^{i m (\phi-\theta(t))} k dk \nonumber \\
& + \frac{1}{2 \pi} \sum_{m = -\infty}^{\infty} \int_{0}^{\infty} \frac{1}{A^3} \left \{ k \ddot{\rho}(t) {J'_{|m|}} (k \rho(t)) + {k}^2 (\dot{\rho}(t))^2 {J''_{|m|}} (k \rho(t)) - 2 i k m \dot{\rho}(t)\dot{\theta}(t) {J'_{|m|}} (k \rho(t)) \right . \nonumber \\
& \qquad \qquad \qquad \qquad \quad \left . - i m \ddot{\theta}(t) J_{|m|} (k \rho(t)) - m^2 (\dot{\theta}(t))^2 J_{|m|} (k \rho(t)) \right \} J_{|m|} (k r) e^{i m (\phi-\theta(t))} k dk \nonumber \\
& - \frac{1}{2 \pi} \sum_{m = -\infty}^{\infty} \int_{0}^{\infty} \frac{1}{A^4} \left \{ k^3 (\dot{\rho}(t))^3 {J'''_{|m|}} (k \rho(t)) - 3 i k^2 m (\dot{\rho}(t))^2 \dot{\theta}(t) {J''_{|m|}} (k \rho(t))  - 3 k m^2 \dot{\rho}(t) (\dot{\theta}(t))^2 {J'_{|m|}} (k \rho(t)) \right . \nonumber \\
& \qquad \qquad \qquad \qquad \quad \left . + i m^3 (\dot{\theta}(t))^3 J_{|m|} (k \rho(t)) +k \dddot{\rho}(t) J'_{|m|}(k\rho(t)) + 3k^2 \dot{\rho}(t) \ddot{\rho}(t) J''_{|m|}(k\rho(t))  \right . \nonumber \\
& \qquad \qquad \qquad \qquad \quad \left . - 3ikm \ddot{\rho}(t) \dot{\theta}(t) J'_{|m|}(k\rho(t)) - 3ikm \dot{\rho}(t) \ddot{\theta}(t) J'_{|m|}(k\rho(t))  \right . \nonumber \\
 & \qquad \qquad \qquad \qquad \quad \left. - im \dddot{\theta}(t) J_{|m|}(k\rho(t)) - 3m^2 \dot{\theta}(t) \ddot{\theta}(t) J_{|m|}(k\rho(t)) \right \} J_{|m|} (k r) e^{i m (\phi - \theta(t))} k dk. \label{expanded}
\end{align}
\end{widetext}
The first term in Eq.~\eqref{expanded} is calculated as 
\begin{align}
& \frac{1}{2\pi} \sum_{m=-\infty}^{\infty} \int_{0}^{\infty} \frac{J_{|m|} (k \rho)}{k^2 + 1} J_{|m|} (k r) e^{im(\phi - \theta)} k dk \nonumber \\
&= \frac{1}{2\pi} \int_{0}^{\infty} \frac{1}{k^2 + 1} J_0 \left ( k \sqrt{r^2 + \rho^2 - 2 r \rho \cos(\phi-\theta) } \right ) k dk \nonumber \\
&= \frac{1}{2 \pi} \mathcal{K}_0 \left ( \sqrt{r^2 + \rho^2 - 2 r \rho \cos(\phi-\theta) } \right ). \label{formula}
\end{align}
Here we used the formula in Ref.~\cite{Watson} (Eq.~(4) on p.361 and Eq.~(5) on p.425).
By differentiating both sides of Eq.~\eqref{formula}, we obtain other relations between equations in real and wavenumber spaces, and all other terms can be converted into the representation in real space.
Thus, we have the result presented in Eq.~\eqref{conc_single}.

\section{Driving force \label{dri.for.}}

First, we derive Eq.~\eqref{torque3} as follows:
\begin{align}
&\bm{F}(\bm{\ell}) \nonumber \\
=& \lim_{\rho \to +0} \frac {-1}{\pi \rho^2} \int_{0}^{2 \pi} \gamma \left ( c (\bm{\ell} + \rho \bm{e}(\phi)) \right ) \bm{e} (\phi) \rho d\phi \nonumber \\
=& \lim_{\rho \to +0} \frac {-k}{\pi \rho} \int_{0}^{2 \pi} c (\bm{\ell} + \rho \bm{e}(\phi)) \bm{e} (\phi) d\phi \nonumber \\
=& \lim_{\rho \to +0} \frac {-k}{\pi \rho} \int_{0}^{2 \pi} \left [ c (\bm{\ell}) + \rho \nabla c (\bm{\ell}) \cdot \bm{e}(\phi) + \mathcal{O} (\rho^2) \right ] \bm{e} (\phi) d\phi \nonumber \\
=& -k \nabla c (\bm{\ell}). \label{c1}
\end{align}
Since we expand the concentration field in the integrand around the considered point, $\bm{\ell}$, the above-mentioned derivation is valid only for a concentration field without divergence at $\bm{\ell}$.

When we consider the motion of a single camphor disk with no constraint, the driving force per contact area, $\bm{F}_s$, is calculated as follows:
\begin{align}
& \bm{F}_s (\dot{\bm{\ell}}, \ddot{\bm{\ell}}) \nonumber \\
=& \lim_{\rho \to +0} \frac{1}{\pi \rho^2} \int_0^{2\pi} c_s\left( \bm{\ell} +  \rho \bm{e}(\phi) ; \bm{\ell} \right) \bm{e}(\phi) \rho d\phi \nonumber \\
=& \frac{1}{4\pi} \left ( -\gamma_{\mathrm{Euler}} + \log \frac{2}{\rho} \right ) \dot{\bm{\ell}} - \frac{1}{16\pi} \ddot{\bm{\ell}} - \frac{1}{32\pi} \left | \dot{\bm{\ell}} \right |^2 \dot{\bm{\ell}} + \frac{1}{48\pi} \dddot{\bm{\ell}}.
\end{align}
When the motion of the camphor particle is restricted to be along a circle with a radius of $\ell = \mathrm{const}.$, but no constraint on the angular direction of the circle, then the torque per contact area, $\tau_{ii}$, is calculated as follows:
\begin{align}
\tau_{ii} 
=&\bm{\ell}_i \times \bm{F}_s (\dot{\bm{\ell}_i}, \ddot{\bm{\ell}_i}) \nonumber \\
=& \frac{1}{4\pi} \left ( -\gamma_{\mathrm{Euler}} +\log \frac{2}{\rho} \right ) \ell^2 \dot{\theta} - \frac{1}{16\pi} \ell^2 \ddot{\theta} - \frac{1}{32\pi} \ell^4 \dot{\theta}^3 \nonumber \\
&+ \frac{1}{48 \pi} \ell^2 \left ( \dddot{\theta} - \dot{\theta}^3 \right ).
\end{align}

From Eqs.~\eqref{torque3} and \eqref{c1}, $\tau_{ij}$ $(i \neq j)$ is calculated as 
\begin{align}
\tau_{ij} =& - \bm{\ell}_i \times \nabla c_s(\bm{\ell}_i; \bm{\ell}_j) \nonumber \\
=&- \frac{1}{4\pi} \mathcal{K}_0 \left ( 2 \ell \right ) \ell^2 \dot{\theta} + \frac{1}{8\pi} \mathcal{K}_1 \left ( 2 \ell \right ) \ell^3 \ddot{\theta} \nonumber \\
& - \frac{1}{16\pi} \mathcal{K}_1 \left ( 2 \ell \right ) \ell^5 \dot{\theta}^3 \nonumber \\
& - \frac{1}{24\pi} \mathcal{K}_2 (2 \ell) \ell^4 \left ( \dddot{\theta} - \dot{\theta}^3 \right ).
\end{align}


\begin{thebibliography}{99}
\bibitem{note} We used $\rho = 0.1 e^{1/4}$ to compare with the numerical results by the following reason: In the analytical framework in which the source term is a delta function, the force originating from a camphor disk moving at a constant velocity $v \bm{e}_x$ is written as $\bm{F} = [(-\gamma_{\rm Euler} + \log (2/\rho)) v / (4\pi) - (1/(32\pi)) v^3 + \mathcal{O}(v^5)] \bm{e}_x$. On the while, in the framework that camphor molecules are dissolved inside a circular region with a radius of $R$, it can be written as $\bm{F} = [(-\gamma_{\rm Euler} + \log (2/R) - 1/4) v / (4\pi) - (1/(32\pi)) v^3 + \mathcal{O}(v^5)] \bm{e}_x$~\cite{IKN}. Therefore, these two situations correspond to each other by setting $\rho = R \exp(1/4)$.
\end{thebibliography}
\end{document}